\documentclass[useAMS,usenatbib]{mnras}

\usepackage{bm}
\usepackage{natbib}
\usepackage{graphicx}
\usepackage{amssymb, amsmath}

\renewcommand{\tabcolsep}{1mm}

\title[Neutron transfer reactions]{Neutron transfer reactions in
    accreting neutron stars}
\author[A. I. Chugunov]
{A.~I.~Chugunov$^1$\thanks{andr.astro@mail.ioffe.ru}\\
    $^1$Ioffe Institute, St Petersburg, Russia
}

\begin{document}

\date{Accepted 2018 xxxx. Received 2018 xxxx;
        in original form 2018 xxxx}

\pagerange{\pageref{firstpage}--\pageref{lastpage}}
\pubyear{2018}

\maketitle

\label{firstpage}

\begin{abstract}
I suggest a novel type of nuclear reactions in accreting neutron stars - neutron transfer, which is quantum tunneling of weakly bounded neutron from  one nucleus to another. I estimate the rate of this process for fixed nuclei separation and then average the result over realistic distribution of nuclei to get the rate value for astrophysical conditions. The neutron transfer can modify reaction chains in accreting neutron stars, thus affecting their heating and cooling. In particular, it can suppress cooling by URCA pairs of nuclei, which is supposed to be crucial for the hottest neutron stars.
\end{abstract}

\begin{keywords}
    stars: neutron; nuclear reactions; stars: evolution;
    X-rays: binaries
\end{keywords}

\maketitle

\section{Introduction}

Observations of thermal emission from transiently accreting neutron stars (NSs) are widely applied to constrain properties of superdense matter (see \citealt{mdkse18} for recent review).
It is generally believed, that
the energy source for thermal emission
is the heat released by exothermic nuclear reactions, which ignite in the course of burying of  accreted matter under a newly accreted layers (e.g.\ \citealt*{bbr98,ch08}). In the top layers of NS ($\rho\lesssim 10^9$~g\,cm$^{-3}$) initially accreted light nuclei burn out to heavier `ashes' through complicated chains of nuclear reactions via stable or explosive burning  (see, e.g., \citealt{GK17_XrayBursts,mdkse18} for review). In  a deeper layers,  the main driver of nuclear reactions supposed to be compression of matter by increasing hydrostatic pressure. Namely, compression increases the electron chemical potential $\mu_\mathrm{e}$ (all atoms are completely ionized and electrons are degenerate), so it becomes energetically favourable for nuclei to capture electrons at respective thresholds (e.g.\ \citealt{Sato79,HZ90,HZ03,Gupta_etal07,ch08,HZ08,Fantina_etal18_AccrCrust}). These captures are typically doubled because of nucleon pairing and the second one produces the heat (see \citealt{Fantina_etal18_AccrCrust} for up-to-date details). Subsequent electron captures increases the number of neutrons in the nuclei, thus decreasing neutron separation energy $S_\mathrm{n}$, until neutrons start to drip out from nuclei (e.g.\ \citealt{Sato79,HZ90} and \citealt{Chamel_etal15_Drip} for thermodynamically consistent analysis). The electron captures (and reverse beta-decays)
are typically treated as the only allowed nuclear reactions before neutron drip (see, however, \citealt*{gkm08} and respective discussion at the end of Section \ref{Sec_Implic}). They conserve number of nucleons in the nuclei $A$ and nuclear reaction chains  can be considered separately for each $A$, even if the ashes initially were a complicated nuclear mixture (e.g., supplement of  \citealt{Schatz_Nat14}). Below, I consider the region of densities $\rho\gtrsim 10^{9}$~g\,cm$^{-3}$ up to neutron drip and refer to it as the envelope.

In this letter I suggest a novel type of nuclear reaction, absent in the previous models, -- the neutron transfer reactions, consisting in "hopping" (quantum tunneling) of a neutron from weakly bounded state in one (donor, `d') nucleus  to another (acceptor, `a') nucleus.
Neutron transfer reactions are well known in nuclear physics for near-barrier and sub-barrier  energies
(e.g., \citealt{vonOertzen1987,Zagrebaev03};
\citealt*{Zagrebaev_etal07,Karpov_nExch15}; 
\citealt{WeakBound_rew15}) and the multi-nucleon transfer reaction can be used to synthesize previously unexplored  superheavy elements  (\citealt*{Zagrebaev_etal13,Wuenschel_etal18}).
The specific feature of neutron transfer reactions in NS envelope would be that the main contribution to the neutron transfer rate comes from nuclei located at distance $l_\mathrm{pk}$, which can be as large as $100$~fm, strongly exceeding nuclei radii.
As for thermonuclear reactions, this distance is determined by two competing factors: decrease of the tunneling probability with increase of internuclear distance and decrease of number of nuclei, which can approach closer, because of Coulomb barrier.
In some sense, the neutron transfer reactions are similar to a hopping transition between localized states of electrons, which is the basis of hopping conductivity (see e.g., \citealt{HoppingCond}).
Note, as shown by \cite{Zagrebaev_etal07} neutron transfers can also increase fusion rates due to modification of internucleus potential, but I do not discuss this effect here, limiting myself to neutron transfer reactions only, and, specifically, to order-of-magnitude estimates, which demonstrate that neutron transfer is applicable for astrophysics matters.

Namely, I start from calculation of the transfer probability (per unit time) between static nuclei at given distance, following the approximative 
approach by \cite{NucTranfPert_MB85}; \cite{NucTranst_tomatter}. Then I estimate the reaction rate $\lambda$ (number of reactions per one donor nuclei per unit time) for certain plasma conditions by averaging transfer probability  over realistic internuclei separation in plasma. A simple approximating expression is suggested. Surely, the model (especially its nuclear part) should be improved, and at the end of the letter I briefly discuss its further refinement, leaving detailed calculations for the future.

The neutron transfer rate strongly depends on $S_\mathrm{n}$ of the donor nuclei, charges  $Z_\mathrm{d}$ and $Z_\mathrm{a}$ of both donor and acceptor nuclei, temperature $T$, and electron chemical potential $\mu_\mathrm e$ (which determines density $\rho$ for given composition).
Typically, neutron transfer is relevant for donor nuclei with $S_\mathrm{n}\lesssim 3$~MeV, which  may exist in envelopes of accreting NS.
In particular, it can burn out strongest among URCA pairs -- specific pairs of nuclei which can coexist at respective electron chemical potential $\mu_\mathrm{e}$ (density) and cools down the envelope  by neutrino emission associated with cycles of beta-captures/decay (see   \citealt*{Schatz_Nat14,Meisel_etal15_even_URCA_excl} for details).
As discussed by \cite{Diebel_etal16_URCA_Superb,MD17} neutrino emissivity generated by these pairs can affect superburst ignition.
Furthermore, the neutron transfer, as an exothermic reaction, produces additional heat. Finally, it causes interlacing of reaction chains for different $A$ (typically considered as independent), thus affecting nuclear evolution and energy output in all subsequent reactions.

\section{Neutron transfer reactions}

\subsection{Neutron transfer at given separation}\label{Sec_FixSep}
Let me start from estimation of neutron transfer reaction probability (per unit time) $W(\bm l)$ between given donor and acceptor  nuclei,
separated by radius vector $\bm l$, assuming than nuclei are well separated ($l\gg r_\mathrm{a}+\mathrm{d}$, where $r_i$ is radius of respective nucleus ($i=\mathrm{a},\ \mathrm{d}$).  Following  \cite{NucTranst_tomatter}, I apply Fermi golden rule:
\begin{equation}
 W=\frac{2\pi}{\hbar}\left|M\right|^2 \rho
 \approx 10^{22} \left(\frac{M}{1\,\mathrm{MeV}}\right)^2
  \frac{\rho}{1\, \mathrm{MeV}^{-1}}\,\mathrm{s}^{-1},
\end{equation}
where
$\hbar$ is Plank constant, $M$ is matrix element, and $\rho$ is density of final states.

The most complicated problem is to estimate $M$, which, strictly speaking, requires an accurate model of neutron states for the system of donor and acceptor nuclei (e.g.\ \citealt{Zagrebaev_etal07}). However,  in this letter I restrict myself to a simple order of amplitude estimates,
and suppose that the least bound neutron at donor nucleus is at a high lying state $\Psi_\mathrm{d}$, with binding energy $E$ equal to neutron separation energy for this nucleus $S_\mathrm{n}$. The final state $\Psi_\mathrm{a}$ corresponds to excitation in the nucleus, which has accepted the neutron, and has the same binding energy. I suppose that this excited state rapidly relaxes to the ground state, releasing the heat and preventing reverse reaction.
Following  \cite{NucTranst_tomatter,NucTranfPert_MB85}, I present $M$ as an integral over plane $\Sigma$, which separates nuclei
\begin{equation}
M
=\frac{\hbar^2}{2 m_\mathrm{n}}\int_\Sigma \left(\Psi_\mathrm{a}^\ast\nabla \Psi_\mathrm{d}
-\Psi_\mathrm{d}\nabla \Psi_\mathrm{a}^\ast\right)\mathrm d \bm s.
\label{M_gen}
\end{equation}
Here  $m_\mathrm{n}$ -- neutron mass.
It is worth stressing that the integral do not depend on the location of plane $\Sigma$ (it can be even curved surface) if the nuclear potential, which bounds neutrons, is vanishing on this surface.

To calculate this integral I approximate wave functions at large distances
from nuclei `a' and `d' ($r^\prime\gg r_\mathrm{a},\ r\gg r_\mathrm{d}$) as
\begin{eqnarray}
\Psi_\mathrm{a}&\approx& A_\mathrm{a}
\,\frac{r_\mathrm{ a}}{r^\prime}\exp(-\kappa\,r^\prime),\\
\Psi_\mathrm{d}&\approx& A_\mathrm{d}
\frac{r_\mathrm{ d}}{r}\exp(-\kappa\,r).
\end{eqnarray}
Here $\kappa=\sqrt{2m_\mathrm{n}E}/\hbar$ and $r^\prime=\left|\bm r-\bm l\right|$ -- distance from the center of the acceptor nucleus. The normalizing amplitudes are estimated as $A_\mathrm{i}\approx 1/\sqrt{V_\mathrm{i}}$, where $V_i=4\pi r_\mathrm{i}^3/3$
($i=\mathrm{a},\ \mathrm{d}$).
%
In leading order on $1/(\kappa l)$ the integral (\ref{M_gen}) is
\begin{eqnarray}
M&\approx& \frac{3\,\hbar^2}{2\, m_\mathrm{n}}
\frac{1}{\sqrt{r_\mathrm{d}r_\mathrm{a}}}
\frac{\exp(-\kappa l)}{l}
\label{M} \\
&\approx& \frac{0.3 \mathrm{MeV}}{l_{50}}
\exp\left(
     -11.3\,l_{50}\, \sqrt{\frac{E}{\mathrm{MeV}}}
     \right).
\nonumber
\end{eqnarray}
Here and below approximate numerical expression are given for fiducial values $r_\mathrm{d}=r_\mathrm{a}=4$\,fm and $l_{50}=l/50$\,fm.

I estimate the density of final states as
\begin{equation}
 \rho\approx m_\mathrm{n} V_\mathrm{a} \frac{k_\mathrm{n}}{\pi^2\hbar^2}
 \approx 1\,\mathrm{MeV}^{-1},
\end{equation}
where $k_\mathrm{n}=\sqrt{2\,(U_0-E)\,m_\mathrm{n}}/\hbar$ is neutron wave number inside acceptor nucleus ($U_0\sim 50$~MeV is typical depth of the neutron potential).
Note, that if the transfer decreases ground state mass of a nuclear pair, the transferred neutron goes into excited state  (which should be free) and reaction is allowed. In the opposite case, the final states are already occupied by neutrons, so that reaction is prohibited.

Combining the equations above, I come to final expression:
\begin{eqnarray}
W(l)&=&\frac{6\hbar k_\mathrm{n}}{m_\mathrm{n}\,r_\mathrm{d}}\,
   \frac{r^2_\mathrm{a}}{l^2}\exp(-2\kappa l)
    \label{W_fin}
   \\
   &\approx&
 3\times 10^{21} \,\mathrm{s}^{-1}
  \frac{1}  {l_{50}^2}
   \exp\left(
   -22.6\, l_{50}\, \sqrt{\frac{E}{\mathrm{MeV}}}
   \right).
   \nonumber
\end{eqnarray}
%

\subsection{Neutron transfer in NS envelope}
To estimate the rate $\lambda$  of neutron transfer,  from a given donor nucleus to acceptor nuclei of type `a' in NS envelope,
%
%
I
perform volume averaging of $W(\bm l)$ weighted with the microscopic number density of acceptor nuclei
$\mathfrak n_\mathrm{a}(\bm l)$:
\begin{equation}
\lambda=\int W(\bm l) \mathfrak n_\mathrm{a} (\bm l )\mathrm d^3 \bm l.
 \label{R_gen}
\end{equation}
For isotropic $W(\bm l)$ this integral depends exclusively on
 $\mathfrak n_\mathrm{a}(\bm l)$ averaged over direction of $\bm l$,
\begin{equation}
\mathfrak  n_\mathrm a(l)=n_\mathrm a\, g_\mathrm{ad}(l).
\end{equation}
Here $n_\mathrm a$ is macroscopic number density of acceptor nuclei and $g_\mathrm{ad}(l)$ is pair correlation function.
The latter is well studied by Monte-Carlo and molecular dynamic simulations (e.g., \citealt{Itoh_etal79_screenMixt};
\citealt{Itoh_etal03_EnhRes}; 
\citealt*{cdy07_NucFus,cd09_NucFusMix}; \citealt{Whitley_etal15_screening}).
It can be parametrized by mean-force-potential $u_\mathrm{ad}(l)$:
\begin{equation}
  g_\mathrm{ad}(l)=\exp\left(-\Gamma_\mathrm{ad}\left[\frac{a_\mathrm{ad}}{l}-u_\mathrm{ad}(l)\right]\right).
  \label{g_via_u}
\end{equation}
Here $\Gamma_\mathrm{ad}=Z_\mathrm{a}Z_\mathrm{d} e^2/(a_\mathrm{ad}T)$ is Coulomb coupling parameter, $a_\mathrm{ad}=(a_\mathrm{a}+a_\mathrm{d})/2$. Finally, $a_{i}=Z_i^{1/3} a_\mathrm{e}$ ($i=\mathrm{a},\mathrm{d}$) and $a_\mathrm{e}=(3/4\pi n_\mathrm{e})^{1/3}$, where $n_\mathrm{e}$ is electron number density. Note, $a_{i}$ should not be confused with nuclear radius $r_i$.

Accurate fit for $u_\mathrm{ad}(l)$, applicable for whole possible parameter range for envelopes of accreted NSs, was suggested by \cite{cd09_NucFusMix}, but it
requires numerical integration in (\ref{R_gen}).
To obtain simple analytical expression I apply less accurate fit based on \cite{Itoh_etal79_screenMixt,Itoh_etal03_EnhRes}:
\begin{equation}
u_\mathrm{ad}(l)=1.25-\frac{25}{64} \frac{l}{a_\mathrm{ad}} \label{u_Itoh}.
\end{equation}
It allows integration of (\ref{R_gen}) analytically, in analogy with Gamow integration of thermonuclear reaction rates via saddle point approximation:
\begin{equation}
\lambda= 4\pi n_\mathrm{a} l_\mathrm{pk}^3 \sqrt{\frac{\pi l_\mathrm{pk}}{a_\mathrm{ad}\,\Gamma_\mathrm{ad}}} W(l_\mathrm{pk}) g_\mathrm{ad}(l_\mathrm{pk}). \label{nu_fin}
\end{equation}
The main contribution comes from the nuclei separated by
\begin{equation}
l\approx l_\mathrm{pk}=a_\mathrm{ad}/ \sqrt{\frac{25}{64}+2\frac{\kappa\,a_\mathrm{ad}}{\Gamma_\mathrm{ad}}}.
\label{lpk}
\end{equation}
In case of $1\lesssim l_\mathrm{pk}/a_\mathrm{ad}\lesssim 1.5$  Eq.\ (\ref{nu_fin})  provides very good approximation
of numerical integration of Eq.\ (\ref{R_gen}) for more accurate fits of $u_\mathrm{ad}(l)$,  if one applies  the same fit to calculate $g_\mathrm{ad}(l_\mathrm{pk})$.
Strictly speaking, in opposite case a numerical integration in (\ref{R_gen}) is required to got accurate value of the neutron transfer rate.
However, for $1.5\gtrsim l_\mathrm{pk}/a_\mathrm{ad}$ Eq.\ (\ref{u_Itoh}) underestimates $u(r)$, leading to underestimation of the reaction rate by
Eq.\ (\ref{nu_fin}),
but Eq.\ (\ref{nu_fin}) can be applied to check that the transfer rate is fast enough to be relevant.
For the similar reasons, for opposite case   $ l_\mathrm{pk}/a_\mathrm{ad}\lesssim 1$, $\lambda$ is overestimated by (\ref{nu_fin}) and this equation can be applied to exclude significant reaction flow for given neutron transfer reaction.

Finally, I stress that it is not necessary that donor and acceptor nuclei are of different types. Quite on the contrary, neutron transfer between odd-A nuclei of the same type are often energetically favourable and can happen (see below).

\section{Astrophysical implications: burnout of strongest URCA pairs in the envelope and interlacing of $A$ chains} \label{Sec_Implic}

Before discussing of astrophysical implications it should be noted, that neutron transfer reactions depend on the mass model, which determines the key parameters: $S_\mathrm{n}$ for donor nuclei and $Q_\mathrm{tr}$-value (for negative $Q_\mathrm{tr}$ the transfer is forbidden, see Sec.\ \ref{Sec_FixSep}). Below, I apply experimental nuclear masses from  Atomic Mass Evaluation 2016 (AME2016; \citealt{ame16}),%
\footnote{Table was downloaded from https://www-nds.iaea.org/amdc/.}
if they are available. In opposite case I apply following models: AME2016 extrapolated mass,  FRDM2012  by \cite{FRDM12}, and HFB21  by \cite{HFB21} and compare results.%

Other important feature for neutron transfer in multicomponent plasma is the fact that light nuclei are preferred acceptors -- lower charge allows the nuclei to approach each other closer, making neutron tunnelling simpler (see Eq.\ \ref{nu_fin}). The reaction flow can strongly depend on  abundance of low $Z$ nuclei. I leave detailed studies of these effects beyond the scope of this letter, presenting just a few examples of astrophysicaly important neutron transfer reactions.

Let's start from URCA pairs. As shown by \cite{Schatz_Nat14}, they can lead to strong cooling of envelopes  and, if their abundance  is not negligible, shift superburst ignition to the deeper layers of the NS (\citealt{Diebel_etal16_URCA_Superb}). Strong bounds to the abundance of URCA pairs in envelope of  the hottest known transiently accreting NS -- MAXI J0556-332 was obtained by \cite{Diebel_etal15} by analysis of cooling curves this source.
In previous studies,  only $e^{-}$-capture and $\beta$-decay reactions were considered as main drivers of nuclear composition in the envelope before neutron drip (see e.g., \citealt{lau_etal18}). As a consequence, each $A$-chain can be considered independently, and constraints to the abundance of URCA pairs in envelope can be applied directly to the production of of respective $A$ nuclei by bygone nucleosynthesis near the surface (\citealt{MD17}).

\begin{table}
    \caption{Neutron transfer reaction for the strongest URCA pairs}
    \label{Tab_URCA}
        \begin{tabular}{clcll}
            URCA &
            $\mu_\mathrm{e}$

        &
            Neutron transfer & $Q_\mathrm{tr}$
            & $S_\mathrm{n}^\mathrm{d}$
            \\
            pair
            & [MeV]
            & reaction
            & [MeV]
            & [MeV]
            \\    \hline
            $^{29}\mathrm{Mg} - \,\! ^{29}\mathrm{Na}$
            &
            13.7
            &  $2\,^{29}\mathrm{Mg}
            \rightarrow \,\!
            ^{28}\mathrm{Mg}+\,\! ^{30}\mathrm{Mg}$
            & 2.7
            & 3.7
            \\
            $^{31}\mathrm{Al}-\,\! ^{31}\mathrm{Mg}$
            &12.2
            &  $2\,^{31}\mathrm{Mg}
            \rightarrow \,\!
            ^{30}\mathrm{Mg}+\,\! ^{32}\mathrm{Mg}$
            & 3.5
            & 2.3
            \\
            $^{33}\mathrm{Al} -\,\! ^{33}\mathrm{Mg}$
            &
            13.9
            &  $2\,^{33}\mathrm{Mg}
            \rightarrow \,\!
            ^{32}\mathrm{Mg}+\,\! ^{34}\mathrm{Mg}$
            & 2.4
            & 2.3
            \\
            $^{55}\mathrm{Sc} - \,\! ^{55}\mathrm{Ca}$
            &12.3
            &  $2\,^{55}\mathrm{Ca}
            \rightarrow \,\!
            ^{54}\mathrm{Ca}+\,\! ^{56}\mathrm{Ca}$
            & 1.5 
            & 1.3 
            \\
        \end{tabular}
\end{table}

\begin{figure}
    \begin{center}
        \leavevmode
        \includegraphics[width=8.5cm]{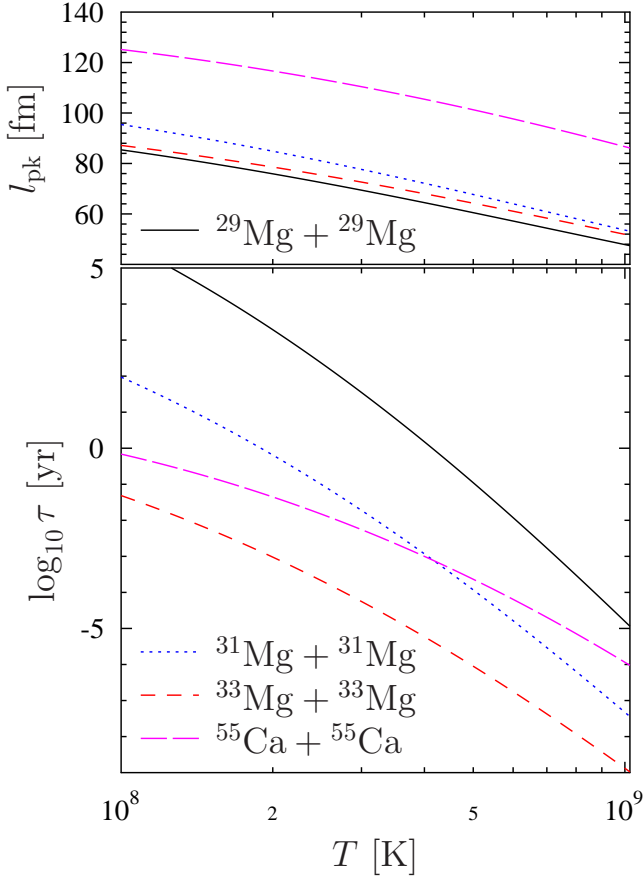}
    \end{center}
    \caption{
        (color online) Timescale of neutron transfer reactions (bottom) and  correspondent $l_\mathrm{pk}$ (top) as function of temperature for reactions, which leads to URCA pairs burnout. For each reaction electron chemical potential is equal to location of correspondent URCA pair (see table \ref{Tab_URCA}). 
    }
    \label{Fig_rate}
\end{figure}

However, if at least one of the elements in the URCA pair has low neutron separation energy, it can turn into a donor for neutron transfer reaction.
Table \ref{Tab_URCA} presents four strongest URCA pairs, listed in the table I by \cite{MD17}, corresponding electron chemical potential,%
\footnote{$\mu_\mathrm{e}$ is given without rest mass. The electrostatic energy  is taken into account;  effects of electron exchange and polarisation, discussed by \cite{cf16}, are neglected for simplicity.}
neutron transfer reaction (between members of the pair), its energy output ($Q_\mathrm{tr}$) and $S_\mathrm{n}$ for donor nuclei.
The reaction timescales $\tau=1/\lambda$ and $l_\mathrm{pk}$ for these reactions  are shown as function of temperature in Fig.\ \ref{Fig_rate} (100\% abundance of the acceptor nuclei is assumed).
Let me note, that reaction $2\,^{31}\mathrm{Mg}
\rightarrow \,\!
^{30}\mathrm{Mg}+\,\! ^{32}\mathrm{Mg}$ also suppresses subsequent URCA pair $^{31}\mathrm{Mg}-^{31}\,\!\mathrm{Na}$, reported by \cite{lau_etal18} as most important for Kepler X-ray burst ashes.
For $^{55}\mathrm{Sc}-\,\!^{55}\mathrm{Ca}$  extrapolated mass from AME2016 is applied.
For FRDM2012  and HFB21 models  $2\,^{55}\mathrm{Ca}
\rightarrow \,\!
^{54}\mathrm{Ca}+\,\! ^{56}\mathrm{Ca}$ rate is suppressed for one and four-five order of magnitude due to different $S_\mathrm{n}$ and $\mu_\mathrm{e}$ values  ($S_\mathrm{n}=1.7$~MeV and $\mu_\mathrm{e}=13.8$~MeV for FRDM2012 and  $S_\mathrm{n}=2.5$~MeV and $\mu_\mathrm{e}=14.8$~MeV for HFB21).

For other pairs from table I by \cite{MD17} the neutron transfer timescale  is too long ($\tau\lesssim 10^{4}$~yr even for $T=7\times 10^8$~K), however, except for $^{57}\mathrm{V}-^{57}\mathrm{Ti}$, all others have low neutrino luminosity.

For applications the timescales shown in Fig.\ \ref{Fig_rate} should be suppressed by the abundance of acceptor nuclei, which is typically rather small $\sim 10^{-3}$ (e.g., table 1 by \citealt{MD17}). However, even after that neutron transfer can be high enough to burnout URCA pair on the accretion timescale  $\tau_\mathrm{acc}\sim 10^{2\div 3}$~yr (e.g.\ Fig. 2 in  \citealt{MD17}).
Note, however, that only for $A=29$ the donor for  neutron transfer is the first member of URCA-pair, i.e.\ neutron transfer can burnout $A=29$ nuclei before respective URCA pair becomes active. For other pairs from table \ref{Tab_URCA} the donor is the second member of the URCA pair, and the neutron transfer starts only than second member is formed by  $\beta$-capture, i.e  URCA pair becomes active. Thus, neutrino emissivity by these URCA pairs can not be prevented, but just suppressed
by gradual decrease of abundance of the pair.

It is worth stressing that the table \ref{Tab_URCA} contains just a limited list of possible neutron transfer reactions with URCA pair members: the member of URCA pair with low separation energy can transfer neutron to any other nuclei in the mixture, if it is energetically favourable. For example, if nuclei with $A=28$ are present in the mixture with mass fraction $\sim 1\%$ (which is  reasonable, see Fig.\ 1 in \citealt{MD17}), the $^{31}$Mg nuclei (generated by e$^-$-capture at $\mu_\mathrm{e}=12.2$~MeV) will participate in neutron transfer reaction $^{28}\mathrm{Mg}+\,\! ^{31}\mathrm{Mg}
\rightarrow \,\!
^{29}\mathrm{Mg}+\,\! ^{30}\mathrm{Mg}$, which leads to burnout of them  on a  timescale of month (for  $T=5\times 10^8$~K). As far as abundance of $A=28$ nuclei is typically larger than for $A=31$ nuclei (see Fig.\ 1 in \citealt{MD17}),  all $A=31$ nuclei can be burned out in this reaction.  Note, this reaction enriches the crust by $^{29}$Mg nuclei, which also have rather low $S_\mathrm{n}\sim 3.7$~MeV and can be donor for neutron transfer reactions (see e.g., first line in Table \ref{Tab_URCA}) or accept neutron from $^{31}\mathrm{Mg}$, if abundance of $^{29}$Mg will become enough large to allow reasonable reaction rate.

It should be stressed that it is not necessary that  donor for neutron transfer reaction belongs to the URCA pair -- the main feature required for the transfer is  low $S_\mathrm{n}$. For example, $^{56}$K, which results from e$^-$ capture by $^{56}$Ca at $\mu_\mathrm{e}\approx 21$~MeV has very low $S_\mathrm{n}$  and can transfer neutron to other nuclei. In particular,
HFB21 model predicts $S_\mathrm{n}\approx 0.3$~MeV for $^{56}\mathrm{K}$ and allows reaction $^{56}\mathrm{K}+\,\! ^{56}\mathrm{Ca}\rightarrow\,\!
^{55}\mathrm{K}+\,\! ^{57}\mathrm{Ca}$, 
with very short timescales ($\sim 10^{-14}$~s) for high  abundance of $^{56}\mathrm{Ca}$. Note, the high abundance of $A=56$ nuclei is realistic (see e.g., Fig.\ 1 in \citealt{MD17}) and one component models typically assume that all nuclei in the crust have $A=56$ before pycnonuclear reactions take place (see, e.g., \citealt{Fantina_etal18_AccrCrust} and references there).
As a result, for HFB21 model the first e$^-$ capture by $^{56}\mathrm{Ca}$ should be followed by rapid neutron transfer to other $^{56}\mathrm{Ca}$ nucleus (instead of second e$^-$ capture, leading to formation of $^{56}\mathrm{Ar}$ in traditional approach).  $^{57}\mathrm{Ca}$, produced by neutron transfer, also captures electron.
The net result of these 3 reactions is
\begin{equation}
2\,\!^{56}\mathrm{Ca}+2\,e^{-}\rightarrow^{55}\mathrm{K}+\,\! ^{57}\mathrm{K}.
\label{3reac}
\end{equation}
It alters all subsequent  reaction chains, since the nuclear composition is different.
Note, the energy output of this reaction is $\sim 35$~keV per nucleon, almost 4 times larger, than energy output
in this layer for traditional approach and the same mass model (see \citealt{Fantina_etal18_AccrCrust}).

However, the reactions followed by $^{56}\mathrm{Ca}+\mathrm{e}^-\rightarrow\,\!^{56}\mathrm{K} $    depend on the mass model. For example, within FRDM2012 mass model $^{56}$K have $S_\mathrm{n}\approx 1$~MeV and cannot transfer neutron to $^{56}\mathrm{Ca}$ ($Q_\mathrm{tr}= -40$~KeV, so thermal activation of neutron transfer reaction is, in principle, possible, but I leave the analysis beyond the scope of this letter). However, the reaction
$^{56}\mathrm{K}+\,\! ^{56}\mathrm{K}\rightarrow\,\!
^{55}\mathrm{K}+\,\! ^{57}\mathrm{K}$ is allowed. For 100\% abundance of $^{56}\mathrm{K}$ the reaction timescale  is   $\sim 2\times 10^{-8}$~s (for $T=5\times 10^8$~K), which exceeds e$^-$-capture rate expected for  $^{56}$K  (see, e.g., \citealt*{Langanke_etal03}). Of course, in realistic reaction network the abundance of $^{56}\mathrm{K}$ will be much smaller, being controlled by its formation rate through e$^-$ capture in $^{56}$Ca and all burn-out channels, including neutron transfer (with $^{56}\mathrm{K}$ or any other nuclei in the same layer as acceptor), but contribution of neutron transfer
should be analysed.

As it was stressed in introduction, the specific feature of the neutron transfer reaction is that  it does not conserve number of nucleons in the nuclei and thus, reaction chains with different $A$ are interlacing and not independent.
Another mechanism of such interlacing is emission
of free neutrons, which can take place after e-capture or as a result of $(\gamma,\ \mathrm n)$ reaction was suggested by \cite{gkm08}.
As an example, let me  take the model, which starts from pure $^{56}$Fe as it was discussed by \cite{lau_etal18}. The two step electron capture by $^{56}$Ca releases neutrons because the most of electron captures at the second step (i.e. electron captures by $^{56}\mathrm{K}$) proceed  to neutron-unbound states of $^{56}\mathrm{Ar}$. The emitted neutrons are rapidly captured by  other nuclei in this layer.
Here I would like to stress two points:
(a) This process takes place in the same layer as neutron transfer reactions, discussed above. Thus realistic model of envelope of accreted neutron star should deal with both types of reactions (neutron transfer and neutron emission from unbound states) simultaneously;
(b) Emission of the free neutrons in the course of two-step electron capture,  depend on the nuclear mass model:  formation of  $^{56}\mathrm{Ar}$ at neutron-unbound state require that energy realise of two step electron capture by $^{56}\mathrm{Ca}$
exceed
separation energy of neutron (or two or more neutrons) for $^{56}\mathrm{Ar}$. It is easy to check, that it is indeed holds true for FRDM mass model, applied by \cite{lau_etal18}.  However,  for family of HFB models by \cite{HFB21,HFB24} it is typically not a case. For example, for HFB21 model the total energy release is just $0.45$~MeV (see table A.1 by \citealt{Fantina_etal18_AccrCrust}), being below neutron separation energy for $^{56}\mathrm{Ar}$ within the same model ($S_\mathrm{n}\approx2.16$~MeV); two neutron separation energy is just a bit smaller $S_\mathrm{2n}\approx1.82$~MeV. Thus, emission of the free neutrons in two step electron capture by $^{56}$Ca is unlikely for HFB21 model.

\section{Summary and caveats}

Summarizing, in this letter I suggest a novel type of neutron transfer reactions, estimate correspondent reaction timescale, and demonstrate that it can affect composition, heating, and cooling of accreting NSs envelopes. Accurate studies of this effects, which account for composition of the nuclear ashes on the top of the envelope, are planned to be performed in subsequent papers.

However, I should note some caveats as well. First of all, more accurate consideration of neutron states, especially in donor nucleus, are crucial to calculate the transfer rate accurately. Such consideration can be done e.g. within Hartree-Fock-Bogoliubov model, which was applied  by \cite{HFB21,HFB24}. In particular, many potential donors nuclei for neutron transfer are strongly deformed, and it can be crucial for transfer rate, but this fact was neglected in this letter. Second, the nuclei in the envelope are not static and their motion can affect the reaction rate (here such effects were neglected in the framework of static approximation).
Third, I discuss neutron transfer from ground state of donor nucleus. However, if donor is formed as a result of electron capture, the least bounded neutron can be in excited state, which can significantly enhance transfer rate.%
\footnote{Within simple model, I consider effect of thermal excitation of neutron states in donor nucleus, but do not got any significant enhancement of transfer rate from nuclei with $S_\mathrm{n}\sim 3$~MeV.}
Finally,  I would like to point that possible effects of neutron transfer on fusion probability (see e.g. \citealt{Zagrebaev_etal07}) should be also discussed with regard to neutron star crust.

\section*{Acknowledgements}
I thank Peter Shternin, Michael Gusakov, Dmitry Georgievich Yakovlev, who have read the draft of manuscript and made useful comments.
I also thank Nicolas Chamel; participants and organizers of `Nucleus' conferences in Voronezh (Russia) in 2012 and 2018 years for fruitful discussions. The last, but not least, I would like to thank Olga Zakutnyaya for her patience, attention and assistance in preparation of the manuscript.


\label{lastpage}
\end{document}